\documentclass[11pt,a4paper]{article}
\usepackage[utf8]{inputenc}
\usepackage{jheppub}
\usepackage{slashed}
\usepackage{amsmath, amssymb, graphicx, physics, dsfont,array,subcaption}
\usepackage[compat=1.1.0]{tikz-feynman}  
\usepackage{comment}
\usepackage{multirow}

\allowdisplaybreaks

\newcommand{\e}{\epsilon}
\renewcommand{\a}{\alpha}
\newcommand{\g}{\gamma}
\newcommand{\G}{\Gamma}
\newcommand{\hyp}{{_2F_1}}

\newcommand{\msb}{{\overline{\text{MS}}}}

\newcommand{\li}[2]{{\text{Li}_{#1}\left(#2\right)}}

\newcounter{RSQ}

\newcounter{MSQ}

\title{Electron and Photon Structure Functions at Two Loops}

\author[a]{Marvin Schnubel}
\author[a]{and Robert Szafron}
\affiliation[a]{
   High Energy Theory Group, Physics Department, Brookhaven National Laboratory, Upton, NY 11973, USA}

\emailAdd{mschnubel@bnl.gov}
\emailAdd{rszafron@bnl.gov}

\abstract{We present a fully analytic computation of the complete electron and photon structure functions, or QED lepton parton distribution functions (PDFs) up to two-loop order. Our computation is performed using modern techniques of reduction to Master Integrals and solving them with the differential equation method.  We obtain explicit expressions for the electron-in-electron, positron-in-electron, photon-in-electron, electron-in-photon, and photon-in-photon distributions at next-to-next-to-leading order (NNLO). Cross-checks against one-loop results, existing two-loop calculations, and a recent soft-collinear effective theory (SCET) analysis of the electron structure functions are presented.
}

\begin{document}
\maketitle
\section{Introduction}
In high-energy scattering processes initiated by electrons (or any charged particle), such as deep inelastic $e$—$p$ scattering or $e^+e^-$ collisions, quantum electrodynamics (QED) corrections often play a dominant role in comparing theory and experiment. These corrections are universal and can be considered independently of specific processes due to the underlying factorization separating long and short distance dynamics. The well-studied QCD collinear factorization \cite{Collins:1985ue,Collins:1989gx} and the concept of a parton  distribution function (PDF) can be naturally extended to leptons in QED. This approach, known as the structure function method \cite{Kuraev:1985hb,Cacciari:1992pz}, provides a systematic way to incorporate universal QED initial state radiation (ISR) corrections~\cite{Frixione:2012wtz,Bertone:2019hks,Bertone:2022ktl,Arbuzov:2024tac}. It has been verified by explicit computation up to the two-loop level \cite{Blumlein:2020jrf}. 

Intuitively, the QED PDFs  describe the probability to find a collinear parton (massless, isolated electron, positron, or photon) carrying a fraction of the "physical" massive electron's momentum. It is important to note that the physical electron, technically speaking, is a non-perturbative object enveloped by a radiation field. 
In contrast, just like in the QCD case, the collinear parton-level leptons are not physical particles, but rather theoretical objects introduced for bookkeeping purposes in QFT.

ISR collinear corrections are enhanced by logarithms of the center-of-mass energy relative to the electron mass, $\ln(Q^2/m^2)$\footnote{For composite particles, the mass in the collinear logarithm is replaced by other low energy scales, such as the charge radius.}. 
These large logarithmic corrections can be summed to all orders in perturbation theory by invoking a renormalization group (RG) equation. In QCD, and by analogy in QED, the corresponding RG equation is known as the DGLAP equation \cite{Dokshitzer:1977sg,Gribov:1972ri,Altarelli:1977zs}. 
Unlike QCD PDFs, which are nonperturbative objects, and only their scale evolution is calculable in perturbation theory, the electron and photon PDFs in QED are entirely perturbative.
This key distinction was noted early on and used to predict structure functions for electrons. The summation of leading logarithmic (LL) QED radiative corrections via a structure-function approach was developed by Kuraev and Fadin \cite{Kuraev:1985hb}, and was instrumental in precision studies at LEP \cite{Skrzypek:1992vk,Cacciari:1992pz,Montagna:1996jv}. 

An alternative approach to ISR proposed by Yennie, Frautschi, and Suura (YFS) \cite{Yennie:1961ad} enables the exponentiation of infrared divergences in QED amplitudes, establishing the foundation for soft-photon resummation techniques. The YFS method effectively resums universal infrared factors, although it does not explicitly provide partonic distributions and does not distinguish collinear and soft scales, which is essential in modern effective field theory frameworks \cite{ Bauer:2000yr,Bauer:2001yt, Beneke:2002ph,Beneke:2002ni}.   
The widespread adoption of YFS has led to QED and QCD  taking drastically different paths during the LEP and LHC eras. However, at very high energies or when very high precision is required, it is advantageous to treat QED and QCD in a unified coherent framework. This, coupled with the pursuit of precision at future lepton colliders, has led to a recent resurgence of interest in research on QED PDFs of leptons. 

Our work follows the collinear factorization approach, focusing on obtaining analytic forms of the electron and photon PDFs at the two-loop order. Previous one-loop (NLO) and two-loop (NNLO)  PDFs provide benchmarks for our calculation. We validated the existing results and demonstrated that modern techniques of differential equations allow for an automated and efficient evaluation of the structure functions needed for future colliders.  

While the studies in the literature often focus on the electron PDFs, here we consider both photon and electron functions on the same footing. Photon PDFs are highly relevant for proposed $\gamma-\gamma$ and $e-\gamma$ colliders \cite{Bogacz:2012fs,Asner:2001vh,Gessner:2025acq,ECFADESYPhotonColliderWorkingGroup:2001ikq,Ginzburg:1982yr, Telnov:2007pe,DeRoeck:2003cjp}. Further, even $e^+ e^-$ and to a lesser extent $\mu^+ \mu^-$ muon collider beams contain photon components \cite{Reuter:626396}.  These appear due to bremsstrahlung effects induced by beam-beam interactions and simulated in  classical electrodynamics with the help of numerical codes such as GUINEA PIG \cite{Rimbault:2007wfy} or WrapX \cite{Fedeli:2022oxl}. 
Secondary photons are particularly relevant in the proposed Wake-Field colliders \cite{Gessner:2025acq}
and hence, a complete study of physics potential must include beam spectra together with lepton and photon PDFs.

In this article, we compute the complete set of NNLO QED PDFs relevant for precision computations needed for planned future colliders \cite{Altmann:2025feg, FCC:2025lpp, Gessner:2025acq, LinearColliderVision:2025hlt}. Our approach relies on solving the two-loop integrals in $x$-space (momentum fraction space) directly via the differential equations method \cite{Henn:2013pwa,Lee:2012cn,Laporta:2000dsw}. The first NNLO evaluation of the electron PDFs has been obtained in Mellin space in \cite{Blumlein:2011mi,Ablinger:2020qvo}, and it is confirmed by our result. The NLO photon PDFs have been validated against the results of \cite{Frixione:2019lga}. The NNLO photon PDFs, $f_{i,\gamma}$, $i=e,\overline{e},\gamma$ can be extracted from heavy quark in gluon PDfs~\cite{Buza:1995ie,Buza:1996wv,Bierenbaum:2007qe,Bierenbaum:2009zt}, by appropriate choice of color factors. Our explicit computation verifies these results and further serves as a cross-check of the massive collinear factorization at the operatorial level.

This paper is organized as follows: we set out our conventions and renormalization scheme in Sec.~\ref{sec:conventions}. 
Then we discuss the technical details of the computation in Sec.~\ref{sec:DiffEq}, focusing on the set-up of integration by parts (IBP) identities and the differential equation method for the master integrals relevant to two-loop PDF calculations. Our  
analytic results are discussed in Sec.~\ref{sec:Results}. We discuss consistency checks and comparisons with existing results in Sec.~\ref{sec:checks}.  Finally, Sec.~\ref{sec:concl} contains our conclusions and future directions.  We provide all master integrals needed in our computation in the App.~\ref{app:MIs}.
In addition, the arXiv version of this article includes an ancillary file, containing all structure functions up to two-loop order.

\paragraph{Note added} During the final stages of this work Ref. \cite{Stahlhofen:2025hqd} appeared in which the two-loop contributions to the channels $f_{e,e}$, $f_{\bar{e},e}$ and $f_{\gamma,e}$ were computed directly in the $x$-space, including the effects of differently flavored leptons in the loops. In this article, we also present the two-loop computation of $f_{e,e}$, $f_{\bar{e},e}$ and $f_{\gamma,e}$ which agree with the expressions given in \cite{Stahlhofen:2025hqd} and furthermore include the photon channels $f_{e,\gamma}$ and $f_{\gamma,\gamma}$. Since we use a different and independent approach to the computation by using a  reduction to Master Integrals which are then solved through differential equations, this serves as a non-trivial cross check of the results. We have adjusted the text in this work to reflect this discussion.

\section{Conventions and Renormalization\label{sec:conventions}}
We use light-cone coordinates throughout this work: for any vector $v^\mu$, we define $v^+ = n\cdot v$ and $v^- = \overline{n}\cdot v$ with $\overline{n}^\mu = (1,0,0,1)$ and $n^\mu = (1,0,0,-1)$, as well as transverse components $v_T$. We note that $n^2 =\overline{n}^2 =0$ and $\overline{n}\cdot n = 2$.  
\subsection{QED Basics }
We calculate the PDFs in QED, in the light-cone gauge. The bare Lagrangian is given by
\begin{align}
\mathcal{L}_0
&= -\frac{1}{4}\,F_{0\,\mu\nu} F_0^{\mu\nu}
   + \bar{\psi}_0\!\left(i \slashed{\partial} - m_0\right)\!\psi_0
   - e_0\,\bar{\psi}_0 \slashed{A}_0 \psi_0
   \;-\; \frac{1}{2\xi}\,\big(n\!\cdot\!A_0\big)^2 \,.
\label{eq:QEDbareLC}
\end{align}
The fields with index zero denote the bare fields and are related to the renormalized fields via
\begin{align}
\psi_0 = \sqrt{Z_2}\,\psi, \qquad  A_{0} = \sqrt{Z_3}\,A \,.
\end{align}
The electric charge $e$ and mass $m$ are renormalized according to 
\begin{align}
e_0 = Z_e\, e \qquad  m_0 = Z_m\, m \,.
\end{align}
 We also use  $\alpha = e^2/(4\pi)$ as the fine-structure constant. We note that in QED Ward identities enforce 
\begin{equation}
  Z_e = \frac{1}{\sqrt{Z_3}} ,
\end{equation}
within any renormalization scheme chosen for the photon sector.

We choose to renormalize the electron and photon wavefunction as well as the charge in the $\msb$ scheme, and the mass in the on-shell scheme. The renormalization factors then read up to two-loop order
\begin{align}
    Z_2^\msb=&1-\frac\a{4\pi\e}+\left(\frac\a{4\pi}\right)^2\left(\frac{11}{6\e^2}+\frac{7}{4\e}\right)+\mathcal{O}(\a^3)\,,\notag\\
    Z_3^\msb=&1+\frac\a{3\pi\e}+\left(\frac\a{4\pi}\right)^2\frac2\e+\mathcal{O}(\a^3)\,,\notag\\
    Z_m^\text{OS} =& 1-\frac{3\a}{4\pi}\left(\frac1\e+\ln\frac{\mu^2}{m^2}+\frac43\right)+\left(\frac\a\pi\right)^2\bigg[\frac{13}{32\e^2}+\frac1\e\left(\frac{73}{64}+\frac{13}{16}\ln\frac{\mu^2}{m^2}\right)+\frac{13}{16}\ln^2\frac{\mu^2}{m^2}\notag\\
&+\frac{73}{32}\ln\frac{\mu^2}{m^2}+\frac{475}{128}-\frac{23\pi^2}{48}+\frac{\pi^2}2\ln(2)-\frac34\zeta_3\bigg]+\mathcal{O}(\a^3)  \,.
\end{align}
To preserve the Ward identity, $Z_e$ and $Z_3$ must necessarily be treated in the same scheme. Note that the $\msb$ $Z$ factors only remove UV poles, while in the on-shell scheme spurious IR poles remain. Additionally, in the $\msb$ scheme the operators have non-trivial LSZ factors. For the photon wavefunction, the LSZ factor $R_\g^\msb$ is defined as
\begin{equation}
    R_\g^\msb Z_e^\msb= Z_3^\text{OS}\,,
\end{equation}
with $Z_3^\text{OS}$ the wavefunction renormlization factor in the on-shell scheme. At the two-loop level, this factor is given by \cite{Melnikov:2000zc,Broadhurst:1991fy}
\begin{align}
Z_3^\text{OS} =& 1-\frac{2\a}{3\pi}\left(\frac1\e+\ln\frac{\mu^2}{m^2}\right)+\left(\frac\a{4\pi}\right)^2\bigg[\frac{16}{9\e^2}+\frac1\e\left(-4+\frac{32}{9}\ln\frac{\mu^2}{m^2}\right)+\frac{16}9\ln^2\frac{\mu^2}{m^2}\notag\\
&-4\ln\frac{\mu^2}{m^2}-30\bigg]+\mathcal{O}(\a^3)\,.\label{eq:Z3OS}
\end{align}
We also will need the well-known running of the QED coupling at one-loop level that is governed by the RGE
\begin{equation}
    \frac{d\a}{d\ln(\mu)}=\beta(\a)=\frac23\frac{\a^2}{\pi}+\mathcal{O}(\a^3)\,.
\end{equation}

\subsection{Gauge choice}
We quantize QED in the light-cone gauge, defined by the gauge-fixing condition $n\cdot A = 0$, where $n^\mu$ is a light-like four-vector. The photon propagator in light-cone gauge is 
\begin{equation}
D^{\mu\nu}(k) = \frac{-i g^{\mu\nu} + i\,\frac{n^\mu k^\nu + n^\nu k^\mu}{n\cdot k}}{k^2 + i0^+}\,,
\end{equation}
This propagator has spurious singularities when $n\cdot k \to 0$. We do not adopt  the conventional Mandelstam-Leibbrandt prescription \cite{Mandelstam:1982cb, Leibbrandt:1983pj} as the dependence on $n\cdot k$ is analytic  (see also \cite{Becher:2010pd} for additional details). An important benefit of light-cone gauge in our context is that the Wilson lines that appear in the definition of PDFs  become unity (see Sec.~\ref{sec:PDF_ops}).

\subsection{Operator Definitions of PDFs }\label{sec:PDF_ops}
The PDFs in QED can be defined through (Fourier-transformed) gauge-invariant bilocal operators analogous to those in QCD \cite{Collins:1981uw}\footnote{We could equivalently use EFT formulation based on SCET \cite{Beneke:2010da}}:
\begin{align}
    \mathcal{O}_{e}(x) &= \frac{1}{2}\int \frac{d\xi^-}{2\pi}e^{-i x P^+ \xi^-}\, \bar{\psi}(\xi^-) \gamma^+ W[\xi^-,0] \psi(0)\,, \\
   \mathcal{O}_{\gamma}(x) &= \frac{1}{P^+}\int \frac{d\xi^-}{2\pi}\, e^{-i x P^+ \xi^-}\, F^{+}_{\;\;\alpha}(\xi^-) F^{+\alpha}(0)\,,
\end{align}
and analogously for the positron operator. $W[\xi^-,0]$ denotes the path-ordered Wilson line (gauge link) connecting the points $0$ and $\xi^-$ along the light-cone: 
\begin{equation}
W[\xi^-,0] = \exp\!\left(-ie\int_0^{\xi^-} dz^- A^+(z^-)\right)\,.
\end{equation}
This Wilson line is required to maintain $U(1)_{\rm em}$ gauge invariance of the nonlocal fermion operator. As the abelian field strength tensor is already gauge invariant, the Wilson line does not appear for the photon operator. 
In light-cone gauge $A^+ = 0$, we have $W[\xi^-,0] = 1$. This is a major simplification: we can ignore the gauge link in our calculations (up to issues of the $1/(n\cdot k)$ pole prescription in loop integrals, which we handle as mentioned).

To calculate the PDFs,  we must evaluate matrix elements of  these operators between external states of physical particles. These states can be related, via the LSZ reduction formula, to amplitudes for appropriate scattering processes, which we compute perturbatively. In practice, we consider the forward matrix element of a time-ordered product consisting of the PDF operator and fields that create/annihilate the external particle. LSZ reduction dictates that each external on-shell leg contributes a factor of $\sqrt{R_e}$ ($\sqrt{R_\gamma}$) for each external electron (photon) field.

For an unpolarized electron (photon) carrying momentum $P^\mu$, the bare electron PDFs: $f_{e,e}(x,\mu)$, ($f_{e,\gamma}(x,\mu)$); and photon PDFs $f_{\gamma,e}(x,\mu)$, ($f_{\gamma,\gamma}(x,\mu)$) are given by matrix elements:
\begin{align}
f_{k,e}(x,\mu) &=  \langle e(P)|\mathcal{O}_{k}(x) |e(P)\rangle_{\mu}\,, \label{eq:opdef_ek}\\
f_{k,\gamma}(x,\mu) &= \langle \gamma(P)|\mathcal{O}_{k}(x) |\gamma(P)\rangle_{\mu}\,, \label{eq:opdef_gammak}
\end{align}
where $k\in \{e,\bar{e},\gamma\}$ and we average over the polarization states of external  states\footnote{We adopt the convention that  the external photon polarization is defined in $d=4$ dimensions, i.e. we work in the 't Hooft-Veltman scheme.}. 
The definitions \eqref{eq:opdef_ek} and \eqref{eq:opdef_gammak} are the lepton-analogues of those for parton densities in a hadron. The normalization is such that at leading order $f^{(0)}_{i,j}= \delta_{ij}\delta(1-x)$, where $\delta_{ij}$ is the Kronecker delta.

The PDF operators contain UV divergencies, which can be removed by  renormalization constants $Z_{ij}$ (with $i,j = e,\gamma$) that relate bare PDF operators to renormalized ones:
\begin{align}
    \mathcal{O}_i(x)=&\int\limits_0^1dy\int\limits_0^1dz\,\delta(x-yz)Z_{ij}(y)\mathcal{O}_j(z)\\
    =&\int\limits_x^1\frac{dz}{z}\,Z_{ij}(z)\mathcal{O}_j\left(\frac{x}{z}\right)\,.\label{eq:operatorrenorm}
\end{align}
These $Z$-factors are the same as those appearing in the usual Altarelli-Parisi splitting functions. In the $\msb$ scheme, $Z_{ij}$ contains only poles and no finite parts. We include these counterterms in our calculation such that after $\msb$ subtraction the results for $f_{i,k}(x,\mu)$ are finite as $\epsilon\to 0$ and depend on $\mu$ as the factorization scale.

\section{IBP reduction and differential equations\label{sec:DiffEq}}
In multiloop QCD calculations it has become the standard procedure not to compute the contributing Feynman integrals directly, but instead to reduce the computation to an often smaller set of Master Integrals (MIs). Even though our integral expressions involve a cut, this method is useful here, too. We generate all Feynman diagrams that contribute to the individual channels with \textsc{FeynCalc} \cite{Shtabovenko:2024aum} and project out the respective Dirac structure before performing partial-fraction decomposition. The remaining expression is then an expression of the form $M=\sum_n C_n \tilde I_n$ where $C_n$ are rational coefficients in the kinematic invariants $x$ and $n\cdot p$ and the dimensional regulator $\epsilon$, and $\tilde I_n$ are scalar Feynman integrals. Subsequently, we use the \textsc{Mathematica} code \textsc{LiteRed} \cite{Lee:2012cn,Lee:2013mka} to reduce the set of scalar integrals to the set of Master integrals. This is achieved through integration-by-parts (IBP) relations \cite{Tkachov:1981wb,Chetyrkin:1981qh}. While \textsc{LiteRed} uses a heuristic approach to finding these relations, many other reduction codes employ Laporta's algorithm \cite{Laporta:2000dsw,Smirnov:2008iw,Maierhofer:2017gsa,Guan:2024byi,vonManteuffel:2012np}. A critical advantage of \textsc{LiteRed} is its capability of dealing with linear propagators as well as cut propagators, too. This is in fact crucial, since in our computation eikonal propagators are ubiquitous due to the adoption of the lightcone gauge. Naively applying this method to our case is however dangerous, because the algorithm can reduce a finite integral to a combination of rapidity-divergent ones. We account for this effect by choosing a momentum routing that ensures that all resulting MIs are finite. We use inverse unitarity
\begin{equation}
    \delta(n\cdot k)\to i\left(\frac{1}{n\cdot k+i\epsilon}-\frac1{n\cdot k-i\epsilon}\right)
\end{equation}
to rewrite the Dirac delta constraint on the large momentum component as an additional cut propagator that can then be handled by the reduction algorithm, further signifying the need for an algorithm that is capable of handling linear propagators. For all channels combined we find five contributing topologies that can be divided into two sets, depending on whether the external particle is massive or massless. We collect all topologies and the corresponding MIs in appendix \ref{app:MIs}. Each MI $I_n(x,\epsilon)$ of a given integral family $I$ depends analytically on the momentum fraction $x$ and the dimensional regulator $\epsilon$. To solve these MIs we set up a set of (coupled) differential equations, where the derivative is taken with respect to the momentum fraction $x$. The resulting system has the schematic form
\begin{equation}\label{eq:diffEQ}
\frac{d}{dx} \vec{I}(x,\epsilon) = M(x,\epsilon)\, \vec{I}(x,\epsilon)\,,
\end{equation}
where $\vec{I} = (I_1, I_2, I_3,\ldots)^T$ is the vector of MIs in topology $I$ and $M(x,\epsilon)$ is a matrix of rational functions in $x$ and $\epsilon$. To generate suitable boundary conditions, we use, inspired by \cite{Broggio:2021fnr}, the fact that once the MI is inclusively integrated over $x$, it must reproduce the ordinary massive two-loop integral with the $\delta$-constraint removed. These boundary integrals are often known in the literature from previous computations (see e.g. \cite{Bogner:2017xhp}) or can be further reduced themselves to simpler integrals using standard IBP relations for uncut diagrams. Whenever possible, we keep the expression for the MIs fully analytic, i.e. all orders in the dimensional regulator $\e$. This also ensures that there are no singularities in $x$ once the integration over $x$ is performed to determine the integration constants.

Oftentimes, it is sufficient to know the order-by-order expansion in $\epsilon$ of the MIs and the full analytic structure is not necessary. If the differential equation can be brought into what is known as $\epsilon$-form, this becomes particularly simple. In these cases one can find a rational transformation matrix $T(x,\epsilon)$ such that
\begin{equation}
    \frac{d}{dx} \vec{J}(x,\epsilon) = \epsilon A(x)\, \vec{J}(x,\epsilon)\,,\label{eq:difftrans}
\end{equation}
with $\vec{J}(x,\epsilon)=T(x,\epsilon)\vec{I}(x,\epsilon)$. We use the \textsc{Mathematica} package \textsc{CANONICA} \cite{Meyer:2017joq} as well as the code \textsc{Fuchsia} \cite{Gituliar:2017vzm} which are based on the reduction algorithms developed in \cite{Henn:2013pwa,Lee:2014ioa,Gehrmann:1999as} to achieve this task. The solution to \eqref{eq:difftrans} is then straightforward and given by Chen's iterated integrals \cite{d12486da-8faa-3fcf-8e4c-623331ef7897}
\begin{equation}
\begin{aligned}
    \vec{J}(x,\e)&=\mathds{P}\exp\bigg[\e\int_\g dA\bigg]\vec{J}_0(\e)\\
    &=\left(1+\sum\limits_{n=1}^\infty\e^n P_n(x)\right)\vec{J}_0(\e)\,,
    \end{aligned}\label{eq:chenit}
\end{equation}
with 
\begin{equation}
    P_n(x)=\int\limits_{x_0}^xdy_1\int\limits_{x_0}^{y_{1}}dy_2 \,\dots \int\limits_{x_0}^{y_{n-1}}dy_n\,A(y_1)A(y_2)\dots A(y_n) \,.
\end{equation}
In \eqref{eq:chenit} $\mathds{P}$ denotes a path ordering along the integration path $\g$, $\vec{J}_0(\e)$ is the boundary condition. If the matrix $dA$ in \eqref{eq:chenit} is a logarithmic one-form, i.e.
\begin{equation}
    dA=\sum\limits_n a_n\, d\log(x_n) \,,
\end{equation}
where $x_n$ are linear polynomials in $x$ and $a_n$ constant matrices, it has been shown that the solution to \eqref{eq:chenit} only contains harmonic and/or multiple polylogarithms \cite{Henn:2013pwa,Duhr:2014woa,Bogner:2014mha}.

Careful consideration must however be given to MIs that need to be interpreted in the distributional sense after expanding the result in $\epsilon$. These integrals often involve terms like $x^{-1-\epsilon}$ and must be expanded according to
\begin{equation}
    x^{-1-\epsilon}=\frac{-\delta(x)}{\epsilon}+\sum\limits_{k=0}^{\infty}\frac{(-\epsilon)^k}{k!}\left[\frac{\ln(x)^k}{x}\right]_+.
\end{equation}
The aforementioned trick of transforming the differential equation into $\epsilon$-form is not naively applicable in those cases, and the correct $\epsilon$ dependence must be restored first \cite{Lee:2014ioa,Lee:2017qql,Gituliar:2017vzm}. We found, however, one instance where this method did not lead to the correct result. The reason was that a transformation matrix $T$ could not be found that fully diagonalizes the differential equation such that each transformed function (corresponding to $J$ in \eqref{eq:difftrans}) had a unitary power of $\e$ after restoring the correct exponent of the dimensional regulator. In our specific case, two integrals that formed a coupled system read before transforming back $J=x^\e j_1(x)+x^{-1+\e}j_2(x)$ with $j_{1,2}$ rational functions with no singularities in $x$. Eventually, following the standard procedure would lead to spurious singularities in $1/\e$. We resolved the problem by directly computing the MIs with the residue theorem. Computing the integral over $x$ and comparing with the known literature result for the boundary integral serves as a cross-check. We give a collection of all linearly independent MIs in all five topologies and the two topology sets in appendix \ref{app:MIs}.

\section{Results}\label{sec:Results}
\begin{figure}
\centering
\begin{subfigure}{\textwidth}
    \includegraphics[width=\textwidth]{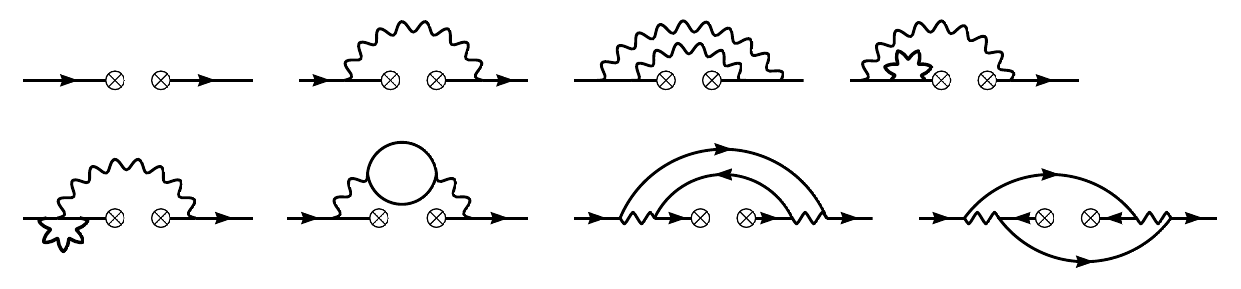}
    \caption{Diagrams contributing to the $e,e$ and $\bar{e},e$ channels at tree (first diagram), one-loop (second diagram) and two-loop level.}
    \label{fig:ee}
\end{subfigure}
\begin{subfigure}{\textwidth}
    \includegraphics[width=\textwidth]{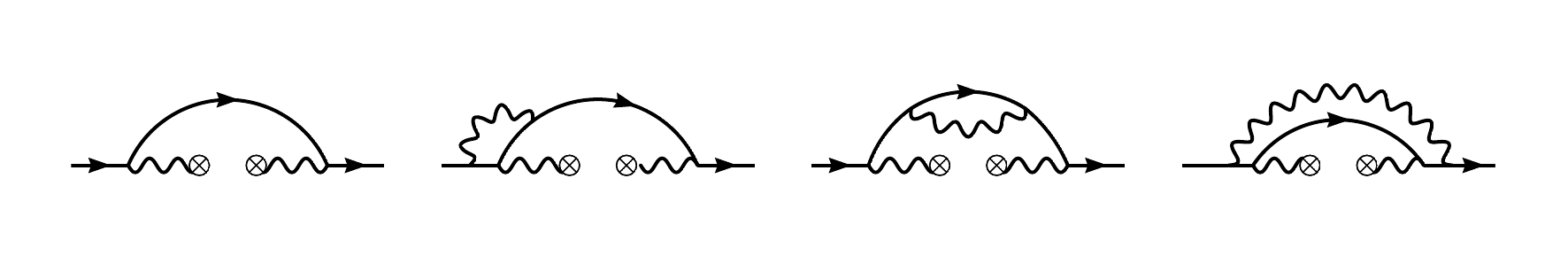}
    \caption{Diagrams contributing to the $\g,e$ channel at one-loop (first diagram) and two-loop level.}
    \label{fig:ge}
\end{subfigure} 
\begin{subfigure}{\textwidth}
    \includegraphics[width=\textwidth]{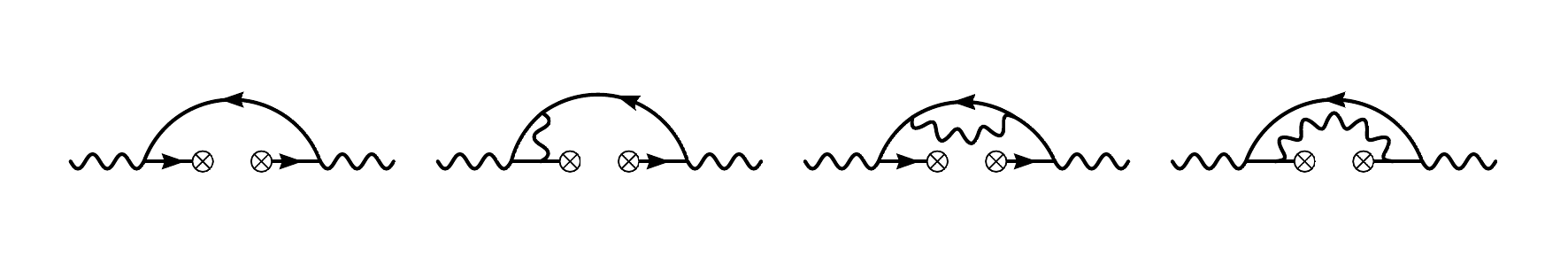}
    \caption{Diagrams contributing to the $e,\g$ channel at one-loop (first diagram) and two-loop level.}
    \label{fig:eg}
\end{subfigure} 
\begin{subfigure}{.75\textwidth}
    \includegraphics[width=\textwidth]{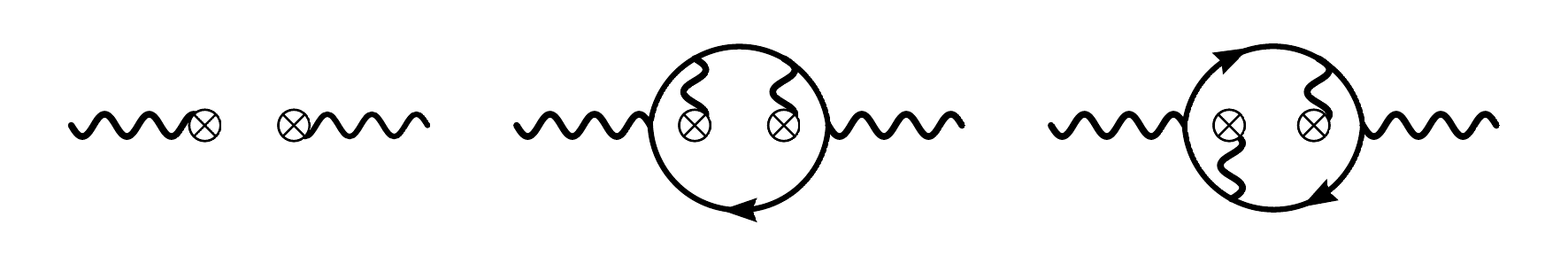}
    \caption{Diagrams contributing to the $\g,\g$ channel at tree (first diagram) and two-loop level. In QED, there is no one-loop contribution.}
    \label{fig:gg}
\end{subfigure} 
\caption{Feynman diagrams contributing to all channels in QED up to two-loop order. Note that in lightcone gauge the collinear Wilson lines (crossed circles) are trivial $W_{\bar{n}}=1$. We omit symmetric diagrams for all channels here.}
\label{fig:diagrams}
\end{figure}

We write the QED structure functions as follows
\begin{equation}
    f_{i,j}(x)=f_{i,j}^{(0)}+\frac\a{2\pi}f_{i,j}^{(1)}(x)+\left(\frac\a{2\pi}\right)^2f_{i,j}^{(2)}\,.
\end{equation}
At tree-level (leading order), the expressions are trivially $f_{e,e}^{(0)}(x)=f_{\g,\g}^{(0)}(x)=\delta(1-x)$, $f_{e,\g}^{(0)}(x)=f_{\g,e}^{(0)}(x))=0$. The NLO results read
\begin{equation}
    \begin{aligned}
        f_{e,e}^{(1)}(x)=&\left[\frac{1+x^2}{1-x}\left(L_m-2\ln(1-x)-1\right)\right]_+ \,,\\
        f_{\g,e}^{(1)}(x)=&\frac{1+(1-x)^2}{x}\left(L_m-2\ln x-1\right) \,,\\
        f_{e,\g}^{(1)}(x)=&(x^2+(1-x)^2)L_m \,,\\
        f_{\g,\g}^{(1)}(x)=&-\frac23L_m \delta(1-x)\,.
    \end{aligned}
\end{equation}
Here, $L_m=\ln\frac{\mu^2}{m^2}$. Even though there does not exist a one-loop QED diagram for the $\g,\g$ channel, the NLO correction is non vanishing due to the LSZ factors needed to calculate the matrix element. These factors amount to the $Z_3$ in the on-shell scheme given in~\eqref{eq:Z3OS}. We give the Feynman diagrams of all channels up to two-loop order in fig. \ref{fig:diagrams}. 
The two-loop results in $x$ space are
    \begin{align}
        f_{e,e}^{(2)}(x)=&L_m^2\bigg[\frac{15-8\pi^2}{24}\delta(1-x)+\frac73L_0(1-x)+4L_1(1-x)-2(1+x)\ln(1-x)\notag\\
        &\quad+\frac{1-5x^2}{2(1-x)}\ln(x)-\frac{2-5x-2x^2-2x^3}{3x}\bigg]\notag\\
        &L_m\bigg[\frac{135-4\pi^2-144\zeta_3}{72}\delta(1-x)+\frac{1+12\pi^2}{9}L_0(1-x)-\frac{34}3L_1(1-x)\notag\\
        &\quad-12L_2(1-x)-\frac72(1+x)\ln^2(x)+6(1+x)\ln^2(1-x)\notag\\
        &\quad+\frac{2(1+x^2)}{1-x}\ln(x)\ln(1-x)-\frac{8+15x-18x^2-12x^3+8x^4}{3x(1-x)}\ln(x)\notag\\
        &\quad+\frac{29+5x}{3}\ln(1-x)+2(1+x)\li{2}{x}\notag\\
        &-\frac{36+(169+18\pi^2)x-(239-18\pi^2)x^2+36x^3}{18x}\bigg]\notag\\
        &+\left(\frac{32077}{2592}+\frac{\pi^2}{4}-\frac{5\pi^4}{36}-2\pi^2\ln(2)+\frac{43}{6}\zeta_3\right)\delta(1-x)\notag\\
        &-\frac{4(13+9\pi^2-108\zeta_3)}{27}L_0(1-x)-\frac{4(3+2\pi^2)}{3}L_1(1-x)+12L_2(1-x)\notag\\
        &+8L_3(1-x)-\frac{1+x}{4}\ln^3(x)-4(1+x)\ln^3(1-x)-\frac{7(1+x^2)}{2(1-x)}\ln^2(x)\ln(1-x)\notag\\
        &+\frac{1+9x^2}{2(1-x)}\ln(x)\ln^2(1-x)-\frac{8-58x+179x^2-208x^3+87x^4+8x^5}{12(1-x)^3}\ln^2(x)\notag\\
        &-10\ln^2(1-x)+\frac{4+29x-38x^2+7x^3+4x^4}{3x(1-x)}\ln(x)\ln(1-x)\notag\\
        &+\frac{8(1+5x-3x^2+2x^3+x^4)}{3x(1-x)}\ln(x)\ln(1+x)\notag\\
        &+\frac{1}{18(1-x)^4(1+x)^3}\big(-115+117x+97x^2+297x^3+351x^4-2329x^5+747x^6\notag\\
        &+1059x^7-440x^8-40x^9\big)\ln(x)-\frac{8-21x+13x^2-(3-5x^2)\pi^2}{3(1-x)}\ln(1-x)\notag\\
        &+\frac{17+5x^2}{1-x}\li{3}{x}+\frac{8(3-x^2)}{1-x}\li{3}{-x}-\frac{1-7x^2}{1-x}\li{3}{1-x}+\bigg(\!\!\!-\frac{11+3x^2}{1-x}\ln(x)\notag\\
        &-\frac{1-7x^2}{1-x}\ln(1-x)+\frac{4+41x-26x^2-2x^3+4x^4}{3x(1-x)}\bigg)\li{2}{x}+\bigg(\!\!-\frac{4(3-x^2)}{1-x}\ln(x)\notag\\
        &+\frac{8(1+5x-3x^2+2x^3+x^4)}{3x(1-x)}\bigg)\li{2}{-x}-\frac{1}{54x(1+x)^2(1-x)^3}\big(112\notag\\
        &-(358+9\pi^2-81\zeta_3)x+(100-39\pi^2)x^2-6(99-8\pi^2+99\zeta_3)x^3\notag\\
        &+4(302+15\pi^2)x^4+(822-59\pi^2+54\zeta_3)x^5-2(238+15\pi^2)x^6\notag\\
        &-2(415-15\pi^2-81\zeta_3)x^7+400x^8\big) \,,
    \end{align}
    \begin{align}
        f_{\bar{e},e}^{(2)}(x)=&L_m^2\bigg[(1+x)\ln(x)+\frac{(1-x)(4+7x+4x^2)}{6x}\bigg]+L_m\bigg[-\frac{2(1+3x+x^2)}{1+x}\ln^2(x)\notag\\
        &\quad-\frac{4(1+x^2)}{1+x}\ln(x)\ln(1+x)-\frac{4(1+x^2)}{1+x}(\li{2}{x}+\li{2}{-x})\notag\\
        &\quad-\frac{(1-x)(8+17x+8x^2)}{3x}\ln(x)-\frac{6+(3+\pi^2)x-12x^2-(3-\pi^2)x^3+6x^4}{3x(1+x)}\bigg]\notag\\
        &-\frac{x}{3(1+x)}\ln^3(x)+\frac{2(1+x^2)}{1+x}\ln^3(1+x)+\frac{1+x^2}{1+x}\ln^2(x)\ln(1+x)\notag\\
        &+\frac{3+84x+290x^2+348x^3+99x^4+8x^5}{12(1+x)^3}\ln^2(x)\notag\\
        &+\frac{4(1+3x+x^2-12x^3-21x^4-7x^5-x^6)}{3x(1+x)^3}\ln(x)\ln(1-x)\notag\\
        &+\frac{8+38x-20x^2-34x^3-8x^4}{3x(1+x)}\ln(x)\ln(1+x)\notag\\
        &-\frac{69+153x+427x^2+335x^3+40x^4}{18(1+x)^2}\ln(x)-\frac{\pi^2(1+x^2)}{1+x}\ln(1+x)\notag\\
        &+\frac{4(1+4x+x^2)}{1+x}\li{3}{x}+\frac{6+32x+6x^2}{1+x}\li{3}{-x}-\frac{12(1+x^2)}{1+x}\li{3}{\frac{1}{1+x}}\notag\\
        &-\bigg(\frac{2(1+8x+x^2)}{1+x}\ln(x)-\frac{8+38x-20x^2-34x^3-8x^4}{3x(1+x)^3}\bigg)(\li{2}{x}+\li{2}{-x})\notag\\
        &+\frac{1}{54x(1+x)^3}\big(-112-(750-45\pi^2-594\zeta_3)x-12(121-7\pi^2-135\zeta_3)x^2\notag\\
        &-18(32-5\pi^2-114\zeta_3)x^3+12(97+9\pi^2+135\zeta_3)x^4+3(442+3\pi^2+198\zeta_3)x^5\notag\\
        &+400x^6\big)\,,
    \end{align}
    \begin{align}
        f_{\g,e}^{(2)}(x)=&L_m^2\bigg[\frac{2-x}{2}\ln(x)+\frac{1+(1-x)^2}{x}\ln(1-x)-\frac{16-28x+11x^2}{12x}\bigg]\notag\\
        &+L_m\bigg[-\frac{2-x}{2}\ln^2(x)-\frac{3(1+(1-x)^2)}{x}\ln^2(1-x)+\frac{32-32x+31x^2}{6x}\ln(x)\notag\\
        &-\frac{20-8x+13x^2}{3x}\ln(1-x)+2(2-x)\li{2}{x}-\frac{32-(5-12\pi^2)x+(85-6\pi^2)x^2}{18x}\bigg]\notag\\
        &+\frac{2-x}{12}\ln^3(x)-\frac{1+(1-x)^2}{6x}\ln^3(1-x)-\frac{4(1+(1-x)^2)}{x}\ln(x)\ln^2(1-x)\notag\\
        &+\frac{12+23x}{24}\ln^2(x)-\frac{96-190x+118x^2-41x^3}{12x^2}\ln^2(1-x)\notag\\
        &+\frac{8(2-x)^3}{3x^2}\ln(1-x)\ln(2-x)-\frac{32-48x+12x^2+7x^3}{6x^2}\ln(x)\ln(1-x)\notag\\
        &-\frac{320-(335+40\pi^2)x+(231+20\pi^2)x^2}{60x}\ln(x)\notag\\
        &+\frac{1}{90x^4}\big(256-640x-400x^2+(1320+120\pi^2)x^3-(1440+120\pi^2)x^4\notag\\
        &+(819+60\pi^2)x^5\big)\ln(1-x)-2(2-x)\li{3}{x}+\frac{8(1+(1-x)^2)}{x}\li{3}{1-x}\notag\\
        &+\frac{16(1+(1-x)^2)}{x}\li{3}{x-1}+\bigg(-(2-x)\ln(x)+4(1-x)\ln(1-x)\notag\\
        &-\frac{32-48x+36x^2-13x^3}{6x^2}\bigg)\li{2}{x}-\frac{8(1+(1-x)^2)}{x}\li{2}{1-x}\notag\\
        &+\bigg(-\frac{8(1+(1-x)^2)}{x}\ln(1-x)+\frac{8(2-x)^3}{3x^2}\bigg)\li{2}{-1+x}\notag\\
        &+\frac{1}{540x^3}\big(1536-192(16-5\pi^2)x+32(41-45\pi^2+135\zeta_3)x^2\notag\\
        &-4(79-180\pi^2+540\zeta_3)x^3-5(401+18\pi^2-216\zeta_3)x^4\big) \,.
    \end{align}
Our findings for the functions $f_{e,e}$, $f_{\bar{e},e}$ and $f_{\g,e}$ agree with those recently given in \cite{Stahlhofen:2025hqd}. The following equations present the two-loop photon structure functions $f_{e,\g}$ and $f_{\g,\g}$.
    \begin{align}
        f_{e,\g}^{(2)}(x)=&L_m^2\bigg[(x^2+(1-x)^2)\ln(1-x)-\frac12(1-2x+4x^2)\ln(x)-\frac{11}{12}+\frac73x-\frac43x^2\bigg]\notag\\
        &+L_m\bigg[(x^2+(1-x)^2)(\ln^2(1-x)-2\ln(x)\ln(1-x))+\frac12(1-2x+4x^2)\ln^2(x)\notag\\
        &+\frac12(3-4x+8x^2)\ln(x)+4x(1-x)\ln(1-x)\notag\\
        &+7-\frac{\pi^2}3-\left(\frac{29}2-\frac{2\pi^2}{3}\right)x+\left(10-\frac{2\pi^2}{3}\right)x^2\bigg]\notag\\
        &+\frac1{12}(1-2x+4x^2)\ln^3(x)+\frac16(x^2+(1-x)^2)(\ln^3(1-x)-3\ln(x)\ln^2(1-x))\notag\\
        &-\frac12(3-6x+2x^2)\ln^2(x)\ln(1-x)-\frac18(1+12x-20x^2)\ln^2(x)\notag\\
        &+\frac12(1+2x-3x^2)\ln^2(1-x)-(1+6x-3x^2)\ln(x)\ln(1-x)\notag\\
        &-\left(2-\frac{\pi^2}{12}+\left(\frac94+\frac{\pi^2}{6}\right)x+\left(6+\frac{\pi^2}{3}\right)x^2\right)\ln(x)\notag\\
        &+\left(\frac{\pi^2}{6}-\left(\frac{13}{2}+\frac{\pi^2}{3}\right)x+\left(6+\frac{\pi^2}{3}\right)x^2\right)\ln(1-x)+(3-6x+2x^2)\li{3}{x}\notag\\
        &+(x^2+(1-x)^2)\li{3}{1-x}-(3-6x+2x^2)\ln(x)\li{2}{x}\notag\\
        &-\left((1-2x)\ln(x)+(x^2+(1-x)^2)\ln(1-x)+\frac12+12x-8x^2\right)\li{2}{1-x}\notag\\
        &+\frac{13}{4}-\frac{\pi^2}{8}-2\zeta_3-\left(\frac{41}{4}-\frac{7\pi^2}{6}-4\zeta_3\right)x+\left(10-\frac{5\pi^2}{6}\right)x^2\,,
    \end{align}
    \begin{align}
        f_{\g,\g}^{(2)}(x)=&L_m^2\bigg[\frac{4}9\delta(1-x)+2(1+x)\ln(x)+\frac{(1-x)(4+7x+4x^2)}{3x}\bigg]\notag\\
        &+L_m\bigg[\!-2(1+x)\ln^2(x)\!-2(3+5x)\ln(x)+\frac{4(1-x)(1-11x-5x^2)}{3x}-\delta(1-x)\bigg]\notag\\
        &+\frac13(1+x)\ln^3(x)+\frac12(3+5x)\ln^2(x)+\left(8+\frac{\pi^2}{3}+\left(12+\frac{\pi^2}{3}\right)x\right)\ln(x)\notag\\
        &-\frac{(1-x)(36-4\pi^2-(324+7\pi^2)x-(108+4\pi^2)x^2)}{18x}-\frac{15}{4}\delta(1-x) \,.
    \end{align}
In these equations we denote with $L_n$ the logarithmic plus distribution according to
\begin{equation}
    L_n(x)=\left[\frac{\ln^n(x)}{x}\right]_+\,.
\end{equation}
The plus distribution is to be understood as
\begin{equation}
    \int dx \,L_n(x)\phi(x)=\int dx\, \frac{\ln^n(x)}{x}\left(\phi(x)-\phi(0)\right) \,,
\end{equation}
with a sufficiently smooth test function $\phi(x)$. $\li{n}{x}$ is the $n$th polylogarithm defined by
\begin{equation}
    \li{1}{x}=-\ln(1-x)\,,\qquad \li{n+1}{x}=\int\limits_0^xdt\,\frac{\li{n}{t}}{t}\,.
\end{equation}

\section{Comparisons and checks}
\label{sec:checks}
As mentioned earlier, the one loop results in $x$-space were  presented in \cite{Frixione:2019lga}, while the one and two loop results in Mellin space for the electron structure functions were known earlier from \cite{Blumlein:2011mi,Ablinger:2020qvo}. Recently, during the final stages of this work, Ref.~\cite{Stahlhofen:2025hqd} appeared computing the two loop structure functions for the $e,e$, $\bar{e},e$ and $\gamma,e$ channels, i.e. those channels where an electron appears in the initial state. 
Additionally, \cite{Stahlhofen:2025hqd} gives the results for those channels when a differently flavored lepton is inserted into the loops. In this work, we omit these additional flavor contributions, but present the structure functions for photon initiated channels, i.e. $e,\g$ and $\g,g$, in addition. Our results in section \ref{sec:Results} agree for the respective channels. Although our results seem to have different functional forms, it is easy to show to be equivalent using standard relations between the polylogarithms.

As an additional check of our computation, the coefficients of the mass logarithms can further be verified through the renormalization group equations of the QED PDFs. The running of the PDFs is governed by their corresponding anomalous dimensions, 
\begin{equation}
    \frac{d}{d\ln(\mu)}f_{i,j}(x)=\int\limits_x^1\frac{dz}{x}\sum\limits_k 2P_{i,k}(z)f_{k,j}\left(\frac{x}{z}\right),
\end{equation}
with $P_{i,j}$ being the respective splitting functions. The splitting functions can be read off from the respective $Z$ factors in \eqref{eq:operatorrenorm}. The two-loop electron splitting functions can be obtained from their QCD counterparts in \cite{Furmanski:1980cm,Ellis:1996nn} through abelianization \cite{Blumlein:2011mi,Ablinger:2020qvo}. The two-loop photon splitting functions in QED were first given in \cite{deFlorian:2016gvk}. With this, we can solve the RGE iteratively. Writing
\begin{equation}
\begin{aligned}
f_{i,j}(x)=&\delta_{ij}\delta(1-x)+\frac{\a}{2\pi}\left(P_{i,j}^{(0)}(x)L_m+F_{i,j}^{(1)}(x)\right)\\
&+\left(\frac\a{2\pi}\right)^2\bigg(A_{i,j}(x)L_m^2+B_{i,j}(x)L_m+F_{i,j}^{(2)}(x)\bigg),
\end{aligned}
\end{equation}
we find for $A_{i,j}$ and $B_{i,j}$
\begin{align}
    A_{i,j}(x)=& \frac12\sum\limits_k P^{(0)}_{i,k}\otimes P_{k,j}^{(0)}\notag\\
    B_{i,j}(x)=&\sum\limits_k\left(P_{i,k}^{(1)}\otimes F_{k,j}^{(0)}+P_{i,k}^{(0)}\otimes F_{i,j}^{(1)}\right).
\end{align}
Here, $P_{i,j}^{(1)}$ are the two-loop splitting functions and the symbol $\otimes$ is understood as the convolution
\begin{equation}
    f(x)\otimes g(x) = \int\limits_x^1\frac{dz}{z} f(z)g\left(\frac{x}{z}\right)\,.
\end{equation}

Sum rules can further be employed to check the viability of our computations. The electron structure functions must satisfy
\begin{equation}
    \int\limits_0^1dx\,x\left[f_{e,k}(x)+f_{\bar{e},k}(x)+f_{\g,k}(x)\right]=1\,,
\end{equation}
with $k=e$,$ \g$. Since $f_{k,k}^{(0)}(x)=\delta(1-x)$, this means that the loop corrections of all channels must sum up to zero, i.e.
\begin{equation}
    \int\limits_0^1dx\,x\left[f_{e,k}^{(n)}(x)+f^{(n)}_{\bar{e},k}(x)+f^{(n)}_{\g,k}(x)\right]=0\,\qquad\text{for }n\neq0\,.
\end{equation}
Given our expressions in the previous section, this can readily be checked, and we have used $f_{e,\g}=f_{\bar{e},\g}$.

\section{Conclusions and outlook}
\label{sec:concl}
We have presented the computation of electron and photon structure functions in QED at the two-loop level, i.e. next-to-next-to-leading order (NNLO) in QED computed directly in momentum space, i.e., as a function of $x$-variable, as opposed to Mellin space approach, which was widely adopted in the early works.  Our results for the electron-initiated channels agree with the computation recently presented in \cite{Stahlhofen:2025hqd} and are consistent with \cite{Ablinger:2020qvo,Blumlein:2011mi}. This constitutes a non-trivial check, because the computations rely on fundamentally different approaches. In \cite{Stahlhofen:2025hqd}, the contributing Feynman diagrams are calculated directly using residue theorems. In our approach, however, we first reduce the contributing diagrams to a set of simpler, scalar Master integrals that are consequently solved using differential equations. The boundary conditions for these equations are then found as ordinary, massive two-loop integrals that are well-documented in the literature. In addition to the electron channels, we also give the two-loop results for the photon-initiated channels. The presentation of the NNLO corrections to the structure functions and the computational technique is the main result of this work.

We have performed multiple consistency checks of our results. We directly confirm the one-loop results of \cite{Frixione:2019lga}. We compared our findings for the two-loop terms to the previous computations carried out in Mellin space \cite{Blumlein:2011mi,Ablinger:2020qvo} and the recent $x$-space computation in SCET \cite{Stahlhofen:2025hqd}. Furthermore, we performed consistency checks of our result against QED sum rules for the structure functions. The coefficients of the mass logarithms were additionally compared to the results obtained through renormalization group running, which invokes the well-known splitting functions as evolution kernels.

The knowledge about the structure functions at higher loop accuracy is a crucial ingredient to precisely describe the collinear physics of initial states for the proposed future electron-positron colliders. Moreover, many future lepton collider concepts can equally well be interpreted effectively as a photon collider. The necessary photon structure functions at the desired accuracy are provided with this work.

While the natural extension to three-loop computation may be feasible, considering the current state-of-the-art in QCD \cite{Ablinger:2022wbb}, phenomenologically, other directions are more pressing. 
In QCD it has been emphasized that polarized PDFs play an equally important role as the unpolarized ones \cite{Nematollahi:2021ynm}. Most importantly, they could have an important impact on the solution of the proton spin puzzle \cite{Cheng:2023kmt}. While no such puzzle exists in QED, polarized PDFs are still an important tool for phenomenological studies \cite{Afanasev:2001zg,Slominski:2004az}, especially in the context of future colliders that might allow for polarized electron beams \cite{Abir:2023fpo}.
Additionally, with the NNLO QED effects at hand, the impact of electroweak corrections should be taken into consideration, too, especially in the context of the discussed 10 TeV partonic center of mass collider \cite{Gessner:2025acq, LinearColliderVision:2025hlt,InternationalMuonCollider:2025sys}.
The computation of these structure functions and effects as well as a thorough study of the phenomenological impact of the two-loop corrections is left for future work.

To further connect our results with the YFS approach, it is natural to extend the analysis to the endpoint region, where soft radiation generates large logarithms. While leading power threshold resummation is understood, the understanding of next-to-leading power (NLP) provides a systematic way to capture the next-to-soft limit and better matching between collinear and YFS approaches. Work in this direction is ongoing, with an explicit treatment of NLP factorization in massive QED \cite{Beneke:2025NLPQED}, based on the NLP SCET framework for resummation.

\acknowledgments We thank Ze Long Liu, Xing Wang, Johann Usovitsch, and Samuel Abreu for useful discussions. 
We are grateful to Johannes Bluemlein and Kay Schoenwald for spotting a typo in the previous version of this article. 
This work is supported by the US Department of Energy under Grant Contract DE-SC0012704. M.S. gratefully acknowledges support from the Alexander von Humboldt Foundation as a Feodor Lynen Fellow.

\appendix
\section{Master integrals}\label{app:MIs}
In this appendix, we give all Master Integrals that appear in the two-loop calculation of QED PDFs. Details on how to derive the set of Master integrals and how they are eventually calculated are given in section \ref{sec:DiffEq}. We can divide the MIs into two sets of topologies, where the first set contains MIs for massive external legs and the second one contains massless initial states. The integral topologies are then given as
\begin{center}
\begin{tabular}{ | m{3.5em} | >{\centering\arraybackslash}m{3.5em} | 
                  m{9em} m{9em} m{9em} | } 
  \hline
  set & topology & \multicolumn{3}{c|}{propagators} \\ 
  \hline\hline
  \multirow{6}{*}{$p^2\!=\!m^2$} 
    & $I^A$ & $D_1$: $k^2-m^2$ & $D_2$: $l^2-m^2$ & $D_3$: $(k+l-p)^2-m^2$\\
    &       & $D_4$: $(l-p)^2$ & $D_5$: $(k-p)^2$ & $D_6$: $n\cdot(l-p)$\\
    \cline{2-5}
    & $I^B$ & $D_1$: $k^2-m^2$ & $D_2$: $l^2-m^2$ & $D_3$: $(k+l-p)^2-m^2$\\
    &       & $D_4$: $(l-p)^2$ & $D_5$: $(k+l)^2$ & $D_6$: $n\cdot(l-p)$\\
    \cline{2-5}
    & $J$   & $D_1$: $k^2$     & $D_2$: $l^2$     & $D_3$: $(k-p)^2-m^2$\\
    &       & $D_4$: $(l-p)^2-m^2$ & $D_5$: $(k+l-p)^2-m^2$ & $D_6$: $n\cdot l$\\
  \hline
  \multirow{4}{*}{$p^2=0$}
    & $K$   & $D_1$: $k^2-m^2$ & $D_2$: $(k-p)^2-m^2$ & $D_3$: $(k+l)^2-m^2$\\
    &       & $D_4$: $(k+l-p)^2-m^2$ & $D_5$: $(l+p)^2$ & $D_6$: $n\cdot(l+p)$\\
    \cline{2-5}
    & $L$   & $D_1$: $k^2$     & $D_2$: $l^2-m^2$ & $D_3$: $(l-p)^2-m^2$\\
    &       & $D_4$: $(l+k)^2-m^2$ & $D_5$: $(k+l-p)^2-m^2$ & $D_6$: $n\cdot l$\\
  \hline\hline
\end{tabular}

\end{center}
Each Master integral can be written as
\begin{equation}
    I_n(a,b,c,d,e,f)=\int\frac{d^Dk}{(2\pi)^D}\int\frac{d^Dl}{(2\pi)^D}\frac{\delta(n\cdot k-x n \cdot p)}{D_1^a \,D_2^b\, D_3^c\,D_4^d\,D_5^e\,D_6^f}
\end{equation}
with $D_i$ denoting the propagators in the respective topology. The Master integrals are then solved with the method of differential equations as explained in Sec.~\ref{sec:DiffEq}. We find
    \begin{align}
        I^A(0,1,1,1,0,0)&=\frac{1}{(4\pi)^4}\frac{m^2}{n\cdot p}\left(\frac{\mu^2}{m^2}e^{\g_E}\right)^{2\e}\G(\e)\G(-1+\e)x^{-2\e}\notag\\
        I^A(1,1,1,0,0,0)&=\frac{1}{(4\pi)^4}\frac{m^2}{n\cdot p}\left(\frac{\mu^2}{m^2}e^{\g_E}\right)^{2\e}\frac{2^{-1+2\e}\sqrt{\pi}}{\G\left(\frac12+\e\right)\G\left(\frac32+\e\right)}\G(-1+2\e)(1+x)^{-4\e}x^{-1+\e}\notag\\
        &\quad\times\bigg[\G(\e)\G\left(\frac32+\e\right)4x\,_2F_1\left(\frac12,2\e,\frac12+\e,\frac{(1-x)^2}{(1+x)^2}\right)\notag\\
        &\qquad+\G(1+\e)\G\left(\frac12+\e\right)(1-x)^2\,_2F_1\left(\frac12,2\e,\frac32+\e,\frac{(1-x)^2}{(1+x)^2}\right)\bigg]\notag\\
        I^A(2,1,1,0,0,0)&=\frac{-1}{(4\pi)^4}\frac1{n\cdot p}\left(\frac{\mu^2}{m^2}e^{\g_E}\right)^{2\e}\frac{2^{-1+2\e}\sqrt{\pi}}{\G\left(\frac32+\e\right)}\G(2\e)\G(1+\e)(1+x)^{-4\e}x^{-1+\e}\notag\\
        &\quad\times(1-x)\,_2F_1\left(\frac12,2\e,\frac32+\e,\frac{(1-x)^2}{(1+x)^2}\right)\notag\\
        I^A(1,1,1,0,1,1)&=C_1(\e)(1-x)^{-1-2\e}+(1-x)^{-1-2\e}I_1(x)\label{eq:masteria}
     \end{align}
    \begin{align}
        I^B(0,1,1,1,1,0)&=\frac{1}{(4\pi)^4}\left(\frac{\mu^2}{m^2}e^{\g_E}\right)^{2\e}\frac{\G^2(\e)}{1-2\e}\frac{2}{1-4^{1-2\e}}(1+x)^{1-4\e}\notag\\
        I^B(1,0,0,1,1,0)&=C_2(\e)x^{-1+\e}(1+x)^{2-4\e}\notag\\
        I^B(1,1,1,0,1,1)&=C_3(\e)(1+x)^{-1-2\e}+(1+x)^{-1-2\e}I_2(x)\notag\\
        I^B(1,0,1,1,1,0)&=(C_4(\e)+H_1(x))\hyp(2\e,2\e,1,-x)\notag\\
        &+(C_5(\e)+H_2(x))\frac{(1+x)^{1-4\e}}{\G(2-4\e)}\hyp(1-2\e,1-2\e,2-4\e,1+x)\notag\\
        I^B(2,0,1,1,1,0)&=\frac{(C_4(\e)+H_1(x))}{m^2}\big(-\e\, \hyp(2\e,2\e,1,-x)+2\e^2 x\, \hyp(1+2\e,1+2\e,2,-x)\big)\notag\\
        &-\frac{(C_5(\e)+H_2(x))}{m^2}\bigg(\frac{\e(1+x)^{1-4\e}}{\G(2-4\e)}\hyp(1-2\e,1-2\e,2-4\e,1+x)\notag\\
        &\qquad-\frac{(1+x)^{-4\e}}{2\G(1-4\e)}\hyp(-2\e,-2\e,1-4\e,1+x)\bigg)\label{eq:masterib}
    \end{align}
    \begin{align}
        J(1,0,1,0,1,0) &=\frac{1}{(4\pi)^4}\frac{m^2}{n\cdot p}\left(\frac{\mu^2}{m^2}e^{\g_E}\right)^{2\e}\G(\e)\G(-1+\e)x^{-2\e}\notag\\
    J(1,1,1,1,0,1)&=\frac{-1}{(4\pi)^4}\frac{1}{(n\cdot p)^2}\left(\frac{\mu^2}{m^2}e^{\g_E}\right)^{2\e}\frac{\G^2(\e)}{1-2\e}x^{-2\e}\notag\\
    J(0,1,0,1,1,0)&=\frac{1}{(4\pi)^4}\frac{m^2}{n\cdot p}\left(\frac{\mu^2}{m^2}e^{\g_E}\right)^{2\e}\G(\e)\G(-1+\e)(1-x)^{-2\e}\notag\\
    J(1,1,0,1,1,0)&=C_6(\e)x^{-2\e}+x^{-2\e}J_1(x)\notag\\
    J(1,1,0,0,1,0)&=m^2\bigg(C_7(\e)x^{-\e}\,_2F_1(\e,-1+2\e,-1+3\e,x)\notag\\
    &\quad+C_8(\e)x^{2-4\e}\,_2F_1(2-2\e,1-\e,3-3\e,x)\bigg)\notag\\
    J(2,1,0,0,1,0)&=C_7(\e)\frac{\e(1-2\e)}{1-3\e}x^{-1-\e}(1-x)\,_2F_1(2\e,1+\e,3\e,x)\notag\\
    &\quad+C_8(\e)\bigg[-\e x^{-4\e}(1-x)\,_2F_1(2-2\e,1-\e,3-3\e,x)\notag\\
    &\qquad+2(1-\e)x^{-4\e}(1-x)\,_2F_1(3-2\e,1-\e,3-3\e,x)\bigg]\label{eq:masterj}
    \end{align}
    \begin{align}
        K(1,0,1,0,1,0)=&m^2\big(C_9(\e)\,_2F_1(2-2\e,\e,2-\e,x)+C_{10}(\e)x^{-1+\e}\,_2F_1(1-\e,-1+2\e,\e,x)\big)\notag\\
        K(2,0,1,0,1,0)=&C_9(\e)\bigg[(2-3\e)(1-x)\,_2F_1(2-2\e,\e,2-\e,x)\notag\\
        &\quad-2(1-\e)(1-x)\,_2F_1(3-2\e,\e,2-\e,x)\bigg]\notag\\
        &+C_{10}(\e)\bigg[(2-3\e)x^{-1+\e}(1-x)\,_2F_1(1-\e,-1+2\e,\e,x)\notag\\
        &\quad-(1-\e)x^{-1+\e}(1-x)\,_2F_1(2-\e,-1+2\e,\e,x)\bigg]\notag\\
        K(1,0,1,1,1,0)=&C_{11}(\e)+K_{1}(x)\notag\\
        K(1,1,1,0,0,0)=&C_{12}(\e)\notag\\
        K(1,1,1,1,0,1)=&C_{12}(\e)\frac{1-\e}{m^2 n\cdot p}\left(\ln(1-x)-\ln x\right)\label{eq:masterk}
    \end{align}
    \begin{align}
        L(0,0,1,1,1,0)=&\frac{1}{(4\pi)^4}\frac{m^2}{n\cdot p} \left(\frac{\mu^2}{m^2}e^{\g_E}\right)^{2\e}\G(\e)\G(-1+\e)\nonumber\\
        L(1,0,1,1,0,0)=&\frac{-1}{(4\pi)^4}\frac{m^2}{n\cdot p} \left(\frac{\mu^2}{m^2}e^{\g_E}\right)^{2\e}\frac{\G^2(\e)}{1-2\e}\,x^{-\e}\label{eq:masterl}
    \end{align}
To obtain the $\e$ expansions to arbitrary order of these MIs we use the package \textsc{HypExp} \cite{Huber:2005yg} to expand the hypergeometric functions. The various integration constants read up to order $\mathcal{O}(\e^0)$
\begin{align}
    C_1(\e)=&\frac{1}{n\cdot p}\left[-\frac{5\pi^2}{6}\frac1\e+\frac{4\pi^2}{3}\ln2-12\zeta_3-\frac{5\pi^2}{3}L_m\right]\notag\\
    C_2(\e)=&\frac{m^2}{n\cdot p}\left[\frac{1}{2\e}+L_m\right]\notag\\
    C_3(\e)=&\frac{1}{n\cdot p}\bigg[\frac{1}{\ln(2)\e^2}\bigg(\frac{\pi^2}{24}+\frac32\ln^2(2)\bigg)+\frac{1}{\ln(2)\e}\bigg(\ln^3(2)+\bigg(\frac{\pi^2}{12}+3\ln^2(2)\bigg)L_m\notag\\
        &+\frac{19}{8}\zeta_3\bigg)+\frac{1}{\ln(2)}\bigg(\frac{41\pi^4}{288}+\frac{5\pi^2}{12}+\frac13\ln^4(2)+\bigg(3\ln^2(2)+\frac{\pi^2}{12}\bigg)L_m^2\notag\\
        &+\bigg(2\ln^3(2)+\frac{19}{4}\zeta_3\bigg)L_m-4\text{Li}_4\left(\frac12\right)-\frac72\ln(2)\zeta_3-\mathfrak{I}\bigg)\bigg]\notag\\
    C_4(\e)=&\frac{1}{(4\pi)^4}\frac{1}{n\cdot p}\left(\frac{1}{2\e^2}+\frac{1}{2\e}(3-2L_m)+L_m^2-3L_m+\frac92+\frac{7\pi^2}{12}\right)\notag\\
    C_5(\e)=&\frac{1}{(4\pi)^4}\frac{1}{n\cdot p}\left(\frac{-1}{\e}+2L_m-3-4\g_E+4\ln2+\frac{2\pi^2}{3}\right)\notag\\
    C_6(\e)=&\frac{1}{(4\pi)^4}\frac{1}{n\cdot p}\bigg[\frac{1}{6\e^2}+\frac{1}{\e}\bigg(-\frac12+\frac13L_m+\frac{\pi^2}{9}\bigg)+\frac13L_m^2+\left(\frac{2\pi^2}{9}-1\right)L_m\notag\\
        &\qquad-\frac83-\frac{\pi^2}{12}+\frac{14}{3}\zeta_3\bigg]\notag\\
    C_7(\e)=&\frac{1}{(4\pi)^4}\frac{1}{n\cdot p}\bigg[\frac{1}{3\e^2}+\frac{1}{\e}\bigg(\frac49+\frac23L_m\bigg)+\frac23L_m^2+\frac89L_m+\frac7{18}+\frac{7\pi^2}{54}\bigg]\notag\\
    C_8(\e)=&\frac{1}{(4\pi)^4}\frac{1}{n\cdot p}\bigg[\frac{7}{3\e^2}+\frac{1}{\e}\bigg(\frac{13}{3}+\frac{14}{3}L_m\bigg)+\frac{14}{3}L_m^2+\frac{26}3L_m+\frac{35}{12}+\frac{11\pi^2}{18}\bigg]\notag\\
    C_9(\e)=&\frac{1}{(4\pi)^4}\frac{1}{n\cdot p}\bigg[-\frac1{\e^2}-\frac1{\e}\left(1+2L_m\right)-2L_m^2-2L_m-1-\frac{\pi^2}{6}\notag\\
        &+\e\left(-\frac43L_m^3-2L_m^2-\left(2+\frac{\pi^2}{3}\right)L_m-1-\frac{\pi^2}{6}\right)\bigg]\notag\\
    C_{10}(\e)=&\frac{1}{(4\pi)^4}\frac{1}{n\cdot p}\bigg[-\frac{1}{2\e}-L_m-2+\e\left(-L_m^2-2L_m-6-\frac{\pi^2}{12}\right)\bigg]\notag\\
    C_{11}(\e)=&\frac{1}{(4\pi)^4}\frac{1}{n\cdot p}\bigg[-\frac{1}{2\e^3}-\frac1{\e^2}L_m+\frac{1}{\e}\left(-11+\frac{5\pi^2}{12}-L_m^2\right)\notag\\
        &-\frac23L_m^3+\left(\frac{5\pi^2}{6}-22\right)L_m-14+\frac{4\pi^2}{3}+\frac{4\zeta_3}{3}\bigg]\notag\\
    C_{12}(\e)=&\frac{1}{(4\pi)^4}\frac{m^2}{n\cdot p}\bigg[\frac{1}{\e^2}+\frac1\e\left(1+2L_m\right)+2L_m^2+2L_m+1+\frac{\pi^2}{6}\bigg]
    \end{align}
The constant $\mathfrak{I}$ in $C_3$ is defined as 
\begin{equation}
    \mathfrak{I}=2\int\limits_0^1d x\,\frac{\ln(x)\li{2}{x}}{1+x}\approx-0.405871
\end{equation}
and eventually drops out of the analytic computation. In \eqref{eq:masteria} to \eqref{eq:masterl} we abbreviated multiple functions that are given as integrals over other MIs. Their explicit expressions are unwieldy long, but possible to obtain fully analytically. We give their definitions below
\begin{align}
        I_1(x)=&\int\limits_1^xdy\bigg\{\frac{1-\e}{m^2\,n\cdot p}\frac{(1-y)^{2\e}}{y}J(1,0,1,0,1,0)(1-y)\nonumber\\
        &-\frac{1-2\e}{2m^2\,n\cdot p}\frac{(1-y)^{2\e}}{y}I^A(1,1,1,0,0,0)(y)\nonumber\\
        &+\frac{1}{2n\cdot p}\frac{(1-y)^{1+2\e}}{y}I^A(2,1,1,0,0,0)(y)\bigg\}\nonumber\\
        I_2(x)=&\int\limits_1^xdy\bigg\{\frac{1-\e}{m^2\,n\cdot p}\frac{(1+y)^{2\e}}{y}I^A(0,1,1,1,0,0)(y)\nonumber\\
        &+\frac{1-2\e}{2m^2\,n\cdot p}\frac{(1-y)(1+y)^{-1+2\e}}{y}I^A(1,1,1,0,0,0)(y)\nonumber\\
        &-\frac1{2\e \,n\cdot p}\frac{(1+y)^{-1+2\e}}{y}(2y+\e-6\e y+\e y^2)I^A(2,1,1,0,0,0)(y)\bigg\}\nonumber\\
        J_1(x)=&\int\limits_1^xdy\bigg\{\frac{1-2\e}{m^2}y^{-1+2\e}J(1,1,0,0,1,0)(y)+y^{2\e}J(2,1,0,0,1,0)(y)\nonumber\\
        &-\frac{1-\e}{m^2}y^{-1+2\e}J(0,1,0,1,1,0)(y)\bigg\}\nonumber\\
        K_1(x)=&\int\limits_1^xdy\bigg\{-\frac{1-2\e}{m^2 y}K(1,0,1,0,1,0)(y)+\frac{1}{y}K(2,0,1,0,1,0)(y)\bigg\}
\end{align}
The functions $H_{1,2}(x)$ appearing in $I^B(1,0,1,1,1,0)$ and $I^B(2,0,1,1,1,0)$ are defined as the integrals
    \begin{align}
        H_1(x)=&-C_2(\e)\frac{(1-2\e)^2}{m^2\G(1-4\e)}\int\limits_1^x\frac{dy}{y}\hyp(1-2\e,-2\e,1-4\e,1+y)\notag\\
        &\times\bigg[4\e^3y(1+y)\hyp(1-2\e,1-2\e,2,-y)\frac{\hyp(1-2\e,1-2\e,2-4\e,1+y)}{\G(2-4\e)}\notag\\
        &\quad-\e(1+y)^{4\e}\hyp(2\e,2\e,1,-y)\frac{\hyp(-2\e,-2\e,1-4\e,1+y)}{\G(1-4\e)}\bigg]^{-1}\notag\\
        H_2(x)=&-C_2(\e)\frac{(1-2\e)^2}{m^2}\int\limits_1^x\frac{dy}{y^{1-\e}}\hyp(2\e,1+2\e,1,-y)\notag\\
        &\times\bigg[4\e^2y(1+y)\hyp(1+2\e,1+2\e,2,-y)\frac{\hyp(1-2\e,1-2\e,2-4\e,1+y)}{\G(2-4\e)}\notag\\
        &\quad-\hyp(2\e,2\e,1,-y)\frac{\hyp(-2\e,-2\e,1-4\e,1+y)}{\G(1-4\e)}\bigg]^{-1}\,.
    \end{align}
Their full analytic expressions can only be obtained order by order in $\e$. Since these are very long, we omit them here. All other MIs that we find are either linearly dependent on the ones given here or are related through simple interchanges $x\leftrightarrow1-x$. Some of the boundary conditions that we used here to determine the integration constants were previously, to the best of our knowledge, only known as an expansion in $\e$. It is therefore instructive to provide their full analytical expression here.
\begin{equation}
\begin{aligned}
   n\cdot p\, (4\pi)^4\int\limits_0^1dx\, I^A(1,1,1,0,1,1)&=\frac{1}{n\cdot p}\left(\frac{\mu^2}{m^2}e^{\g_E}\right)^{2\e}\frac{\G(2\e)}{\e}\bigg[\frac{\pi^2}{6}+5\zeta_3\e+\frac{\pi^4}{30}\e^2\bigg]\\
n\cdot p\,(4\pi)^4\int\limits_0^1d x\,J(1,1,0,0,1,0)&=m^2\left(\frac{\mu^2}{m^2}e^{\g_E}\right)^{2\e}\frac{\G(3\e)\G(1-3\e)\G(1-\e)}{\G^2(2-2\e)}\\
        &\times\bigg[\frac{\G(2\e)\G(1-2\e)\G(3-4\e)}{\G(3-3\e)}\,_2F_1(3-4\e,2-2\e,3-3\e,1)\\
        &+\frac{\G(\e)\G(1-\e)\G(-1+2\e)}{\G(-1+3\e)}\,_2F_1(1-\e,\e,-1+3\e,1)\bigg]\\
        n\cdot p\,(4\pi)^4\int\limits_0^1d x\,J(1,1,0,0,1,0)&=\left(\frac{\mu^2}{m^2}e^{\g_E}\right)^{2\e}\G(\e)\G(1-\e)\G(-1+3\e)\\
        &\times\bigg[\frac{\e\G(1-4\e)}{\G(2-3\e)}\,_2F_1(1-4\e,1-\e,2-3\e,1)\\
        &+\frac{\G(2\e)}{\G(3\e)}\,_2F_1(-\e,2\e,3\e,1)\bigg]
\end{aligned}
\end{equation}
By expanding the last two results in $\e$ we find agreement with the expanded results from the literature \cite{Berends:1997vk,Scharf:1993ds,Bauberger:1994by}.

\bibliographystyle{JHEP}
\bibliography{biblio}

\end{document}